\documentclass[12pt]{elsarticle}
\usepackage[english]{babel}
\usepackage{graphicx}
\usepackage{color}
\usepackage{amsmath}
\usepackage{fixltx2e}
\usepackage{textcomp}
\usepackage{upgreek}

\begin{document}
\author{Micha\l{} Cie\'sla$^1$}
\ead{michal.ciesla@uj.edu.pl}
\author{Jakub Barbasz$^{1,2}$}
\ead{ncbarbas@cyf-kr.edu.pl}

\address{$^1$ M. Smoluchowski Institute of Physics, Jagiellonian University, 30-059 Kraków, Reymonta 4, Poland}
\address{$^2$ Institute of Catalysis and Surface Chemistry, Polish Academy of Sciences, 30-239 Kraków, Niezapominajek 8, Poland.}
%


\title{Modelling of Interacting Dimer Adsorption}

\begin{abstract}Adsorption of dimers is modelled using Random Sequential Adsorption algorithm. The interaction between molecules is given by screened electrostatic potential. The paper focuses on the properties of adsorbed monolayers as well as the dependence of adsorption kinetics on interaction range. We designate random maximal coverage ratios, density autocorrelations and orientational ordering inside layers. Moreover the detailed analysis of adsorption kinetics are presented including discussion of Feder's law validity and new numerical method for modelling diffusion driven adsorption. Results of numerical simulations are compared with experimental data obtained previously for insulin dimers.
\end{abstract}
\maketitle
\section{Introduction}
Since its introduction by Feder \cite{bib:Feder1980}, Random Sequential Adsorption (RSA)  became a well established method used for modelling of adsorption properties. Although at the beginning it was used mainly to model adsorption of simple spherical molecules, recent results shows that it could be effective also for quite complex structures like proteins \cite{bib:Talbot2000, bib:Adamczyk2010, bib:Adamczyk2011}. 
\par
Most of the research effort focuses on adsorption of hard objects where geometry is the only factor affecting properties of obtained monolayers. On the other hand, adsorption is often induced by electrostatic interaction between adsorbate and collector, e.g. \cite{bib:Adamczyk1996}. In such cases, the hard body interaction can still be sufficiently good approximation because the electrostatic forces are screened in a solution which makes them negligible. However in the general case, they should be taken into account to find out the level of systematic error provided by such interactionless approximation. 
\par
The purpose of presented paper is to extend previous investigations of dimers adsorption \cite{bib:Ciesla2012} to include the case on non-negligible electrostatic repulsion. The paper focuses on fundamental properties of dimer monolayers, such us maximal random coverage ratio, density autocorrelation and orientational ordering, as well as on adsorption kinetics. The additional aim is to develop robust numerical procedure to convert data obtained from RSA to values measured during typical adsorption experiment. 
\section{Model}
\begin{figure}[hbt]
  \centering
  \includegraphics[width=1.5in]{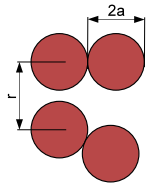}
  \caption{Two dimers at a distance of $r$.}
  \label{fig:dimers}
\end{figure}

A single dimer particle is assumed to consist of two identical, charged, spherical particles (see Fig.\ref{fig:dimers}).
	In RSA studies of soft particles electostatic interaction potential between particles has been typically an exponentially decaying Yukawa potential, due to forming a double layer from solvent particles, which effectively screens the electrostatic charge of an adsorbate. Therefore, such potential is also known as screened electrostatic potential \cite{bib:Senger2000, bib:Trulsson2009}. Here, we assumed that the potential given by a single spherical monomer of radius $a$ is equal to $(U_\text{0}/r) \exp\left[-(r - a)/L_\text{e}\right]$ for $r>a$ and infinite otherwise. Here $r$ denotes distance from a monomer centre. Parameter $U_\text{0}$ characterises electrostatic properties of monomer and of a solvent. Range of the interaction is controlled by $L_\text{e}$ commonly called as the Debye screening length. It is a key parameter for electrostatic interaction in electrolytes and can be calculated as \cite{bib:Debye1923, bib:Adamczyk-book}:
\begin{equation}
L_\text{e}=\sqrt{\frac{\epsilon_\text{0}\epsilon_\text{r} k_\text{B} T}{2e^2 I}}
\end{equation}
	where $\epsilon_\text{0}$ is the permittivity of free space; $\epsilon_\text{r}$ denotes the dielectric constant of a solvent; $k_\text{B}$ is the Boltzmann constant; $T$ denotes temperature; $e$, elementary charge; and $I$, the ionic strength of electrolyte solution. Table \ref{tab:le} contains typical values of $L_\text{e}$ for the most common solutions.
\begin{table}[htb]
\centering
\begin{tabular}{|c|c|c|c|c|}
\hline
Concentration & 1:1       & 1:2 (Na\textsubscript{2}SO\textsubscript{4})  &  2:2            & 1:3 (Na\textsubscript{3}PO\textsubscript{4}) \\
 M                 & (KCl) & 2:1 (CaCl\textsubscript{2})        & (NiSO\textsubscript{4}) & 3:1 (AlCl\textsubscript{3}) \\
\hline
$10^{-1}$ & 0.9639 & 0.5565 & 0.4820 & 0.3935 \\
$10^{-2}$ & 3.048 & 1.759 & 1.524 & 1.244 \\
$10^{-3}$ & 9.639 & 5.565 & 4.819 & 3.935 \\
$10^{-4}$ & 30.48 & 17.59 & 15.24 & 12.44 \\
$10^{-5}$ & 96.39 & 55.65 & 48.19 & 39.35 \\
$10^{-6}$ & 304.8 & 175.9 & 152.40 & 124.4 \\
\hline
\end{tabular}
\caption{$L_\text{e}$ in nm for typical electrolytes characterised by different ratio of cations to anions. Values were taken from \cite{bib:Adamczyk-book}.}
\label{tab:le}
\end{table}
\par
\begin{figure}[hbt]
  \vspace{1cm}
  \centering
  \includegraphics[width=1.5in]{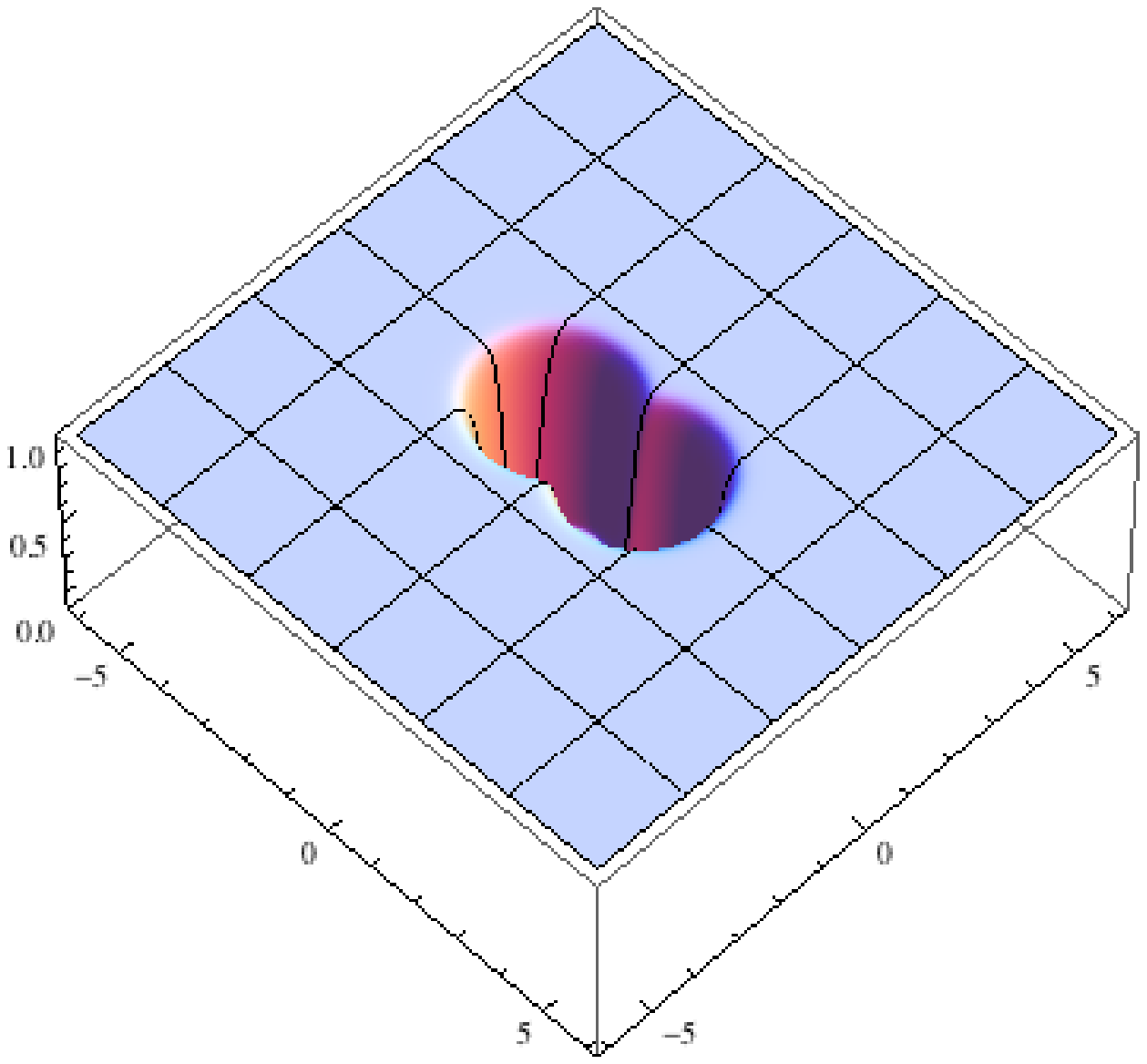}
  \includegraphics[width=1.5in]{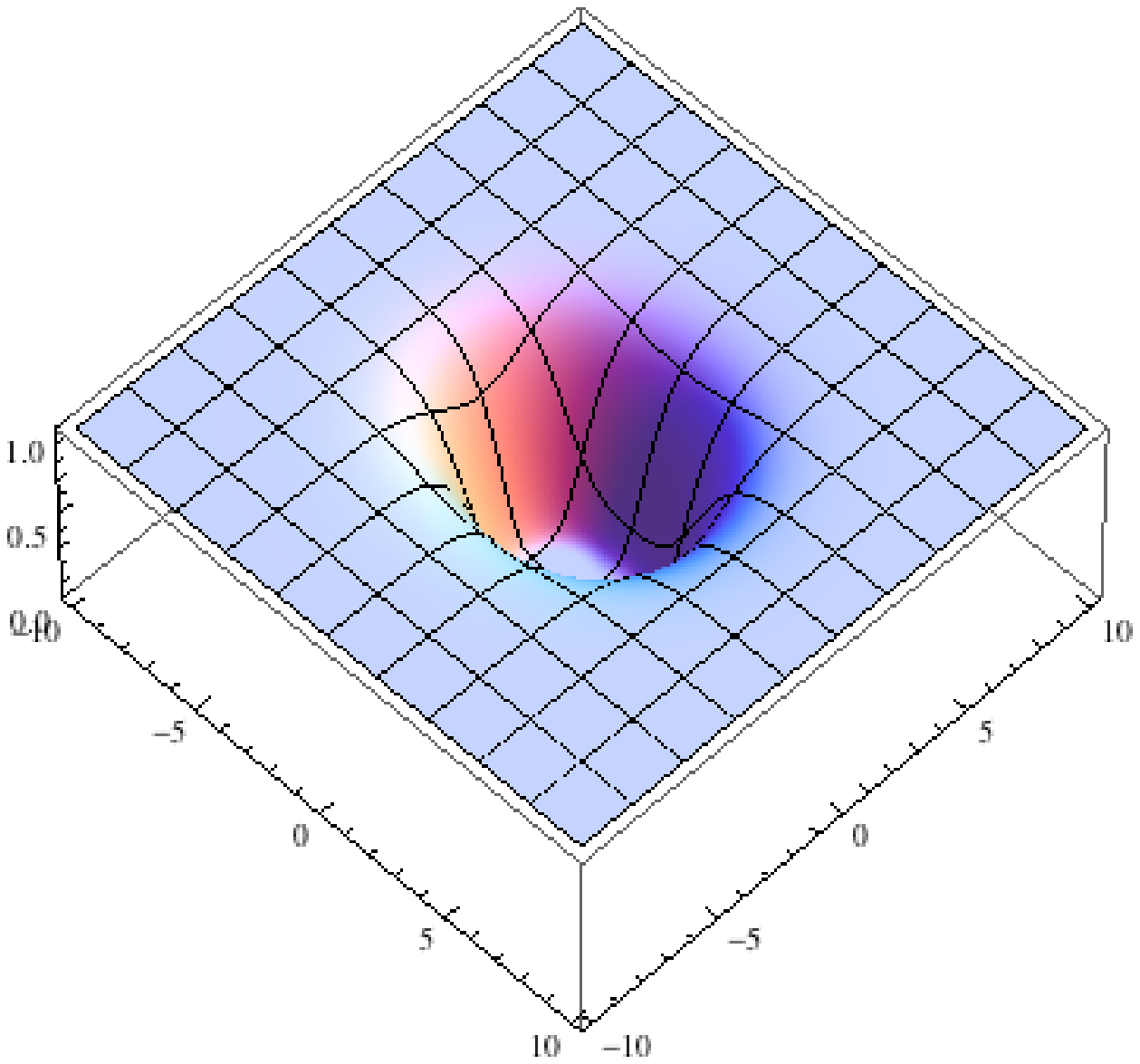}
  \includegraphics[width=1.5in]{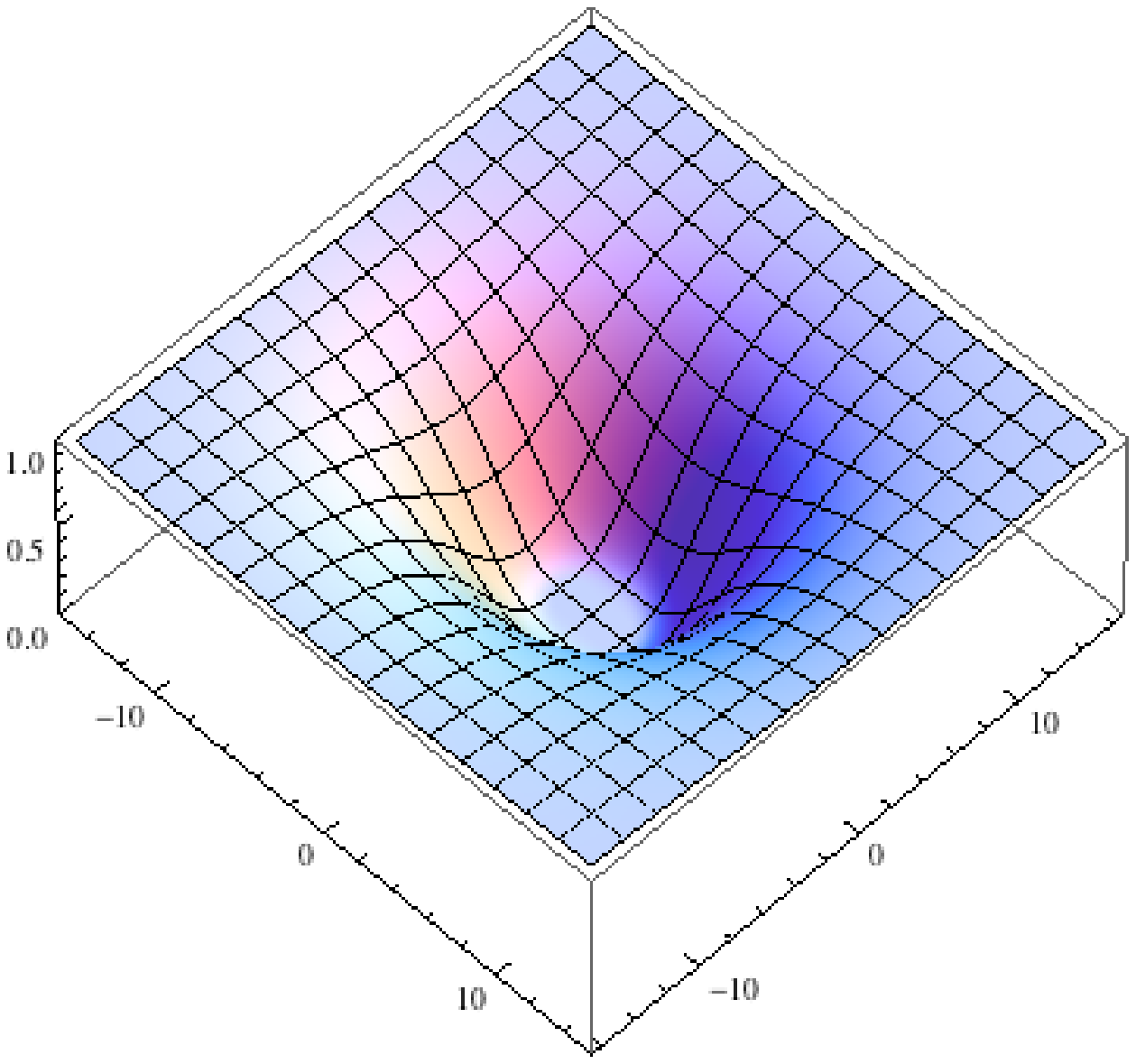}
  \caption{Effective shape of an interacting dimer - probability of adsorption of a point-size repulsive unit charge around a~dimer particle for different Debye screening length $L_\text{e}=0.1\, a$ (left), $1.0\, a$ (centre) and $5.0\, a$ (right). Parameters used for calculations: $U_\text{0}=  6.78 \,\, k_\text{B} T \, a / q^2$, where $k_\text{B}T$, $a$ and $q$ are, respectively, energy, distance and charge units. The grid lines are separated by $2a$. Colors are used for visualization purposes only and do not have any other physical meaning.}
  \label{fig:probability}
\end{figure}

Electrostatic repulsion was introduced to RSA algorithm by Adamczyk et al. \cite{bib:Adamczyk1990} and extended later by Oberholzer et al. \cite{bib:Oberholzer1997}. There, the probability of successful adsorption is assumed to depend on the interaction energy $U$ between the new particle and its nearest neighbour through a Boltzmann factor, $\exp(-U/kT)$, where $U$ includes interaction with both components of a dimer. Therefore, the adsorption probability of a point-like charged particle on a surface with a single dimer will reflect the dimer effective shape, shown in Fig.\ref{fig:probability}.
When point-like particle is substituted by a sphere, the effective potential changes due to different geometry, which changes double layer interactions. In this case:
\begin{equation}
\label{uel}
U_\text{el} = \left\{
  \begin{array}{cl}
  \frac{U_\text{0}}{r}\exp\left[-\frac{r-2a}{L_\text{e}}\right] & \text{for $r\ge 2a$} \\
  \infty & \text{for $r<2a$}
  \end{array}
\right.
\end{equation} 
Note that dimer-to-dimer potential will contain four such terms defining interaction between all pairs of monomers belonging to different particles.
\subsection{Simulation details}
Adsorbed monolayers were generated using modified RSA algorithm. The procedure consists of the following steps:
\begin{description}
\item[I] a new virtual dimer is randomly created. Its centre is set on the collector according to a uniform probability distribution and its orientation (the angle between x-axis and dimer axis) is uniformly chosen from $[0, 2\pi)$;
\item[II.a] the virtual molecule undergoes overlapping test with its nearest neighbours;
\item[II.b] if there is no overlap the total electrostatic potential $U$ between the virtual molecule and previously adsorbed dimers is calculated using Eq.(\ref{uel})
\begin{equation}
U = \sum_{i=1}^2 \sum_{j=1}^{2N} U_\text{el} \left( |\vec{r_i} - \vec{r_j}| \right),
\end{equation}
where $i$ enumerates the virtual particle monomers, $j$ enumerates monomers belonging to previously adsorbed dimers and $\vec{r_i}$ is position of $i$-th monomer centre; $N$ denotes number of already adsorbed dimers.
\item[II.c] a random number is selected according to uniform probability distribution on the interval $[0, 1)$. If it is smaller than $\exp (-U / k_\text{B} T)$ the virtual particle is added to the existing layer.
\item[III] otherwise the virtual dimer is removed and abandoned. 
\end{description}
\par
The whole procedure is repeated for a specified number of times expressed using dimensionless time:
\begin{equation}
t = n \frac{S_\text{m}}{S_\text{c}}
\label{dimt}
\end{equation}
	where $n$ is a number of algorithm iterations, $S_\text{m} = 2\pi a^2$ is a coverage of a single dimer, and $S_\text{c}$ is a collector's surface. 
The fundamental characteristic of an obtained layer is its coverage ratio defined as follows
\begin{equation}
\uptheta = N_\text{m} \, S_\text{m} / S_\text{c}
\label{theta}
\end{equation}
where $N_\text{m}$ is a number of adsorbed particles.
\par
The adsorption process simulation was performed for a squared collector of $200a$-side size and was stopped at $t=10^5$. We did not used periodic boundary conditions, as it had been proved earlier, it does not have a significant influence on obtained layers \cite{bib:Ciesla2012}. For each set of parameters, 20 to 100 independent simulations were performed. Parameters of electrostatic potential (\ref{uel}) were chosen to describe typical experimental conditions of water solutions. Therefore, relative dielectricity of the solvent was $\epsilon_\text{r}=78$. Parameter $a=4.65$ nm provides length scale typical to mid sized bio-molecules. Value of coefficient $U_\text{0} = (e^2a^2)/(4\pi\epsilon_\text{0}\epsilon_\text{r})  = 6.78 \,\, k_\text{B} T$ nm$/e^2$ \cite{bib:Adamczyk-book}, where $\epsilon_\text{0}$ is dielectric constant of vacuum, is fully determined by the above assumptions. The thermal energy $k_\text{B} T$ at a room temperature acts as energy unit.
\section{Results and discussion}
\subsection{Fundamental properties of adsorption monolayer}
Example coverages obtained in numerical simulations are shown in Figure \begin{figure}[hbt]
  \vspace{1cm}
  \centering
  \includegraphics[width=1.5in]{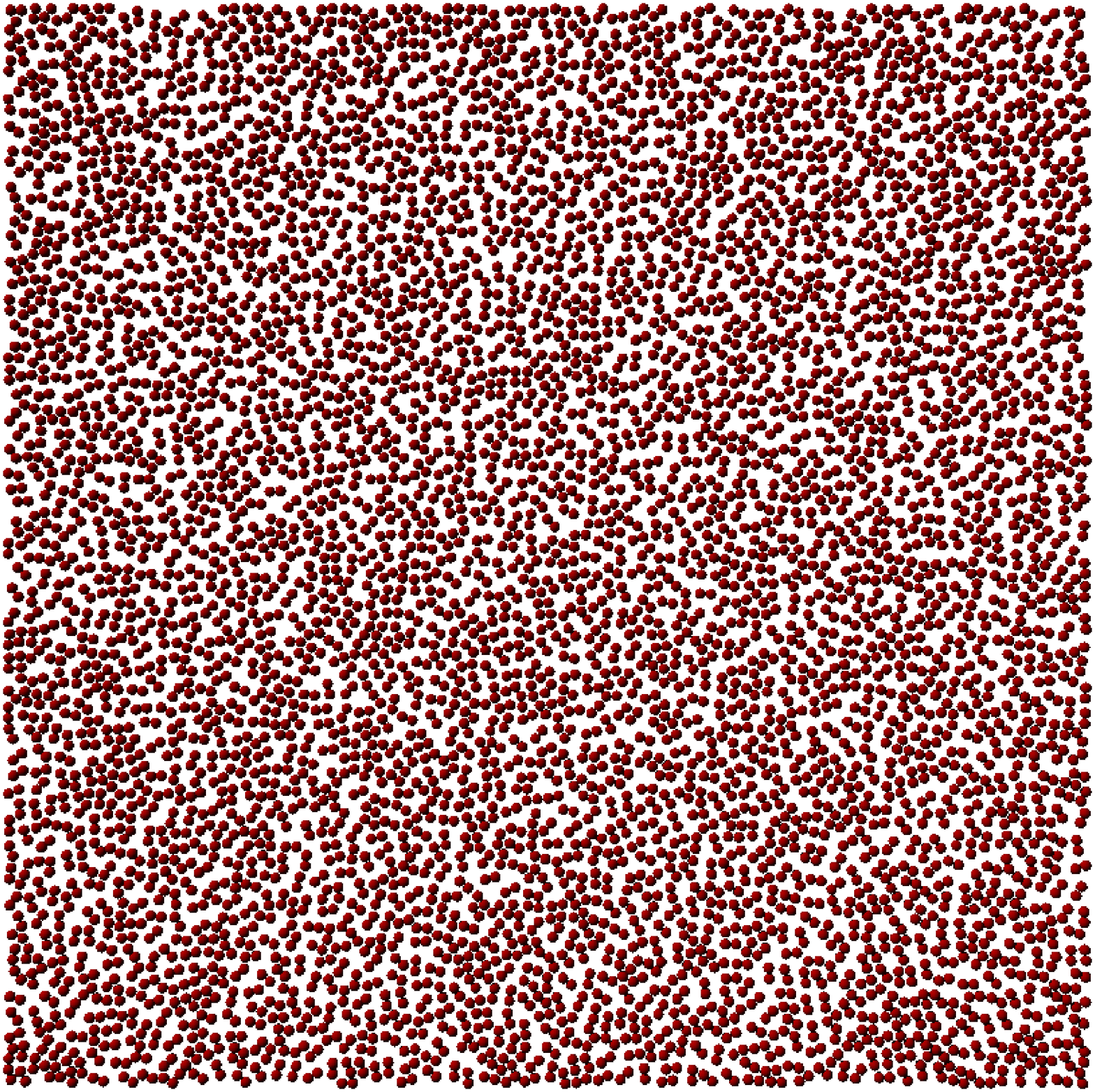}
  \includegraphics[width=1.5in]{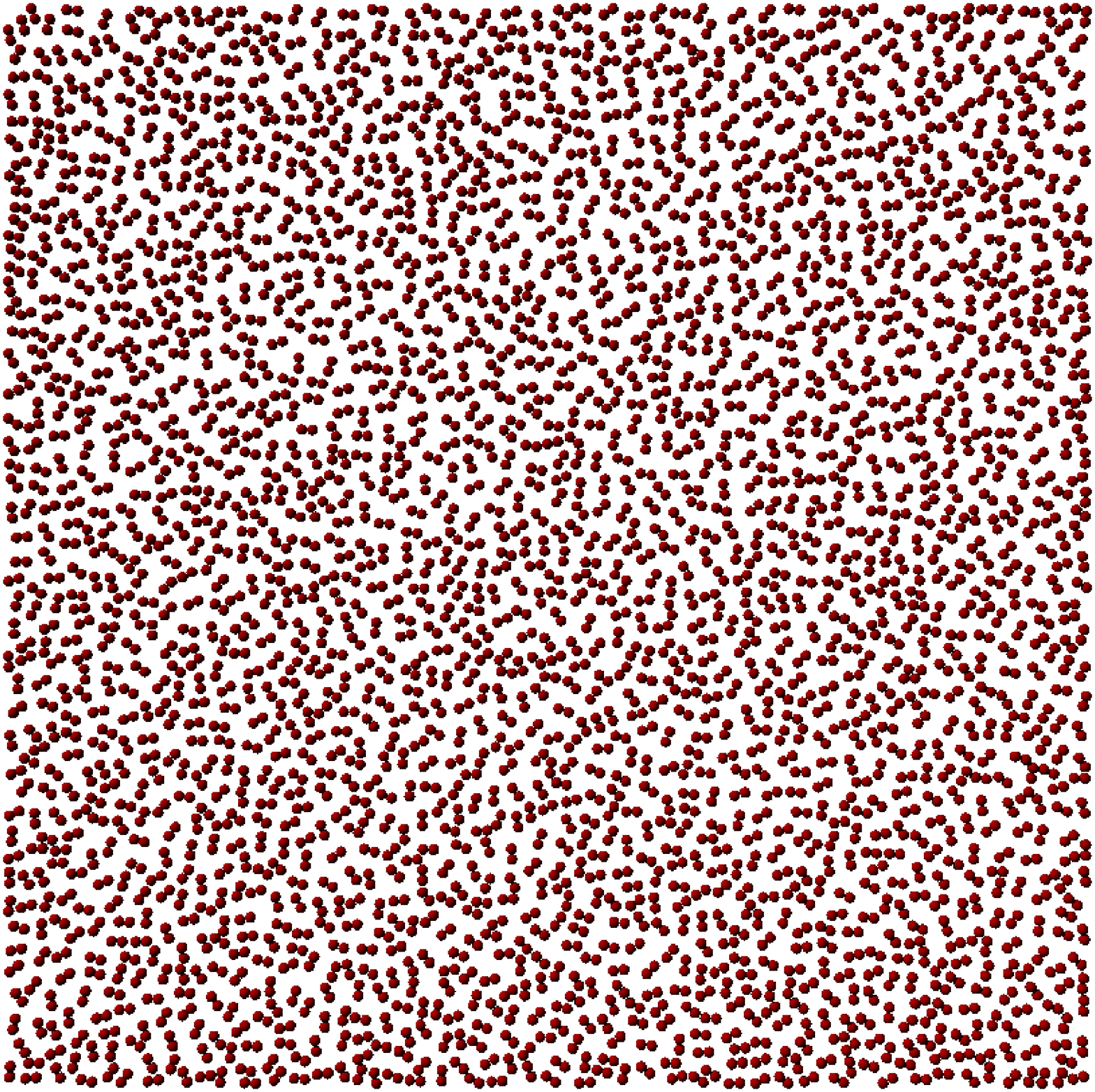}
  \includegraphics[width=1.5in]{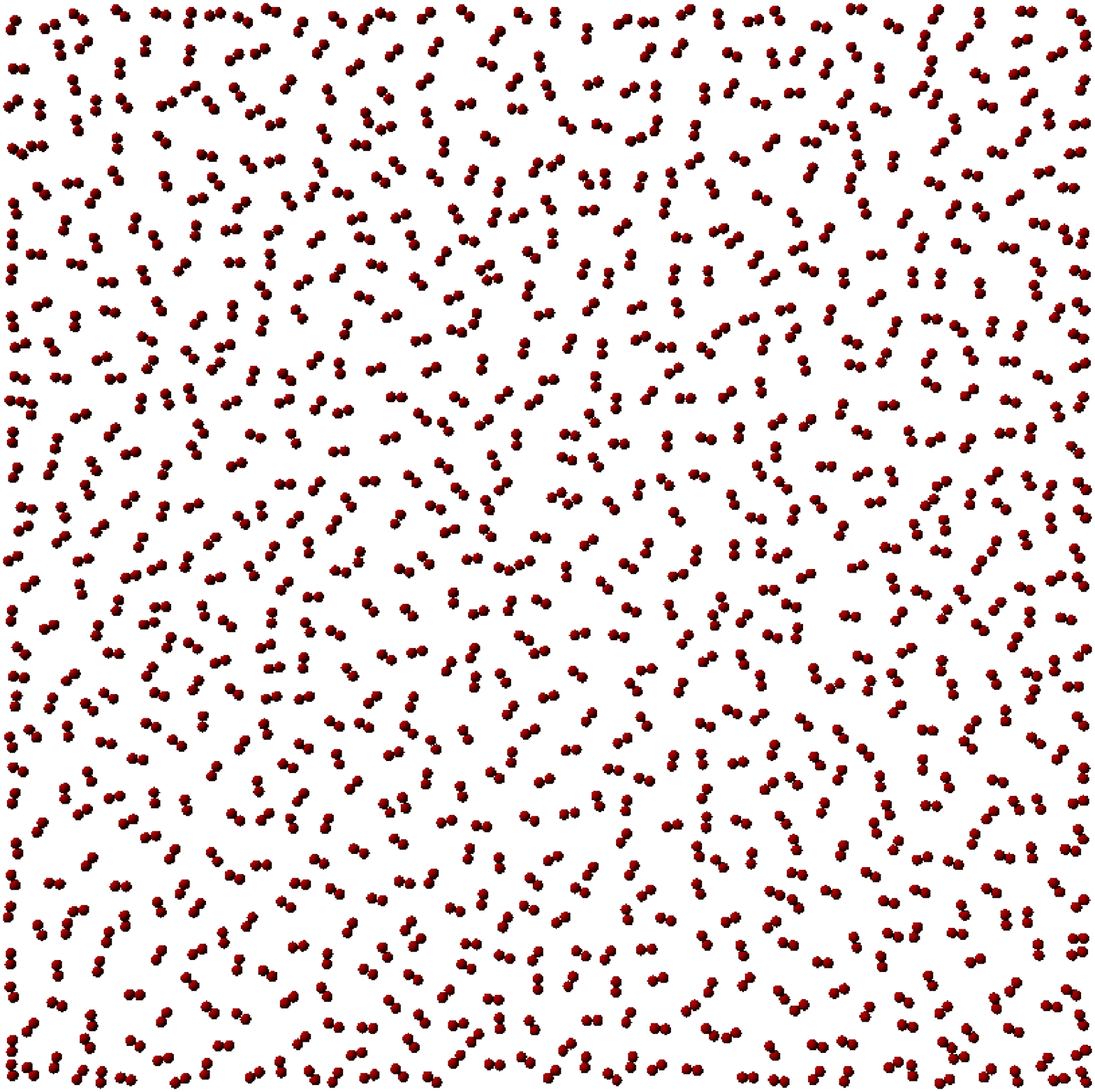}
  \caption{Typical adsorbed layers of dimers for three different electrostatic interaction ranges: $L_\text{e}=0.1 a$, $1.0 a$ and $5.0 a$.}
  \label{fig:coverages}
\end{figure}
Their main properties, such as maximal random coverage ratio, autocorrelation function and orientational ordering, are analysed in the following sections.
\subsubsection{Adsorption ratio}
\par
\begin{figure}[hbt]
  \vspace{1cm}
  \centering
  \includegraphics[width=2.5in]{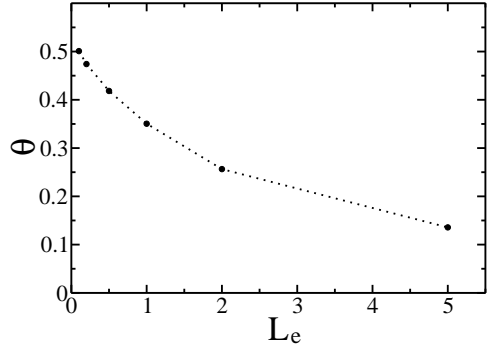}
  \caption{Adsorption ratio dependence on electrostatic interaction range $L_\text{e}$. The ratio is a mean value of 20 to 100 simulations using a collector of $200a$ side size. The time of a single simulation was $10^5$. $L_\text{e}$ is expressed in units of $a$.}
\label{fig:ratio_le}
\end{figure}

The easiest estimation of maximal random coverage ratio can be done by simple counting the number of adsorbed dimers. Figure \ref{fig:ratio_le} shows raw results taken directly from obtained data.
\par
\begin{figure}[hbt]
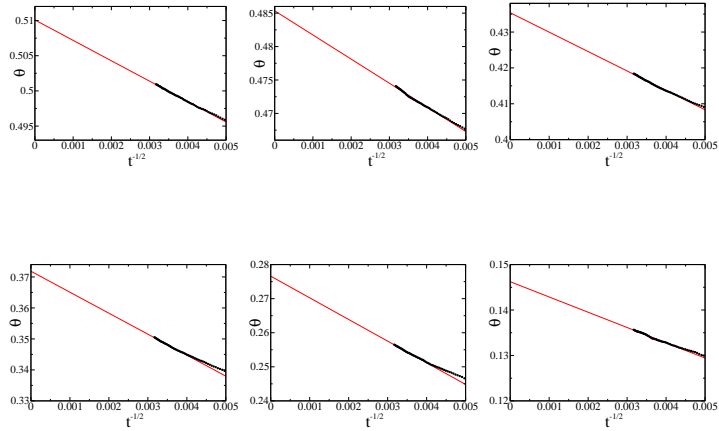

  \centering
  \includegraphics[width=1.2in]{fl01.eps}
  \includegraphics[width=1.2in]{fl02.eps}
  \includegraphics[width=1.2in]{fl05.eps}
  \begin{verbatim}
  \end{verbatim}
  \includegraphics[width=1.2in]{fl10.eps}
  \includegraphics[width=1.2in]{fl20.eps}
  \includegraphics[width=1.2in]{fl50.eps}
	  \caption{Coverage ratio $\uptheta$ dependence on $t^{-1/2}$ for different electrostatic interaction range. $L_\text{e} \in \{0.1,\, 0.2,\, 0.5,\, 1.0,\, 2.0,\, 5.0 \}$. Black dots are taken from simulation and red line is a fit (\ref{federslaw}). The $\uptheta_\text{max}$ is reached when $t^{-1/2}=0$.}
\label{fig:feder}
\end{figure}

However, such ratios are underestimated due to the finite time of a simulation; there is no guarantee that all free places on a collector have been filled. To deal with this, the model of RSA kinetics have to be used. The common choice here is the Feder's law \cite{bib:Swendsen1981, bib:Privman1991, bib:Torquato2006}, which is valid for a wide range of adsorbate molecule's shapes \cite{bib:Viot1992} and also proved to be valid for the case of hard-core dimers RSA \cite{bib:Ciesla2012}:
\begin{equation}
\uptheta_\text{max} - \uptheta(t) = A t^{-1/d}
\label{federslaw}
\end{equation} 
where $t$ is dimensionless time (\ref{dimt}); $d$ is a dimension of a collector and $A$ is a factor of proportionality. Here, $d=2$. Plots illustrating relation (\ref{federslaw}) are presented in Figure \ref{fig:feder}. They confirm that the numerical data obey the Feder's law. Moreover, at the limit of $t \to \infty$ ($t^{-1/2} \to 0$), the maximal random coverage ratios are higher by about 1-2\% than the values obtained directly from data as in Figure \ref{fig:ratio_le} (see Table \ref{tab:ratio_le}).
\par
It is clear that the maximal random coverage ratio defined using (\ref{theta}) and (\ref{federslaw}) decreases with the growth of the Debye screening length $L_\text{e}$ due to the electrostatic repulsion. On the other hand, those results can also be interpreted as an increase in the effective molecule size with $\uptheta_\text{max}=0.547$ being constant.  The effective molecule size can be also estimated analytically: 
\begin{equation}
S_\text{eff}(L_\text{e}) = \int d^2 r \left\{
  \begin{array}{ll}
    1- \exp \left[ -U_\text{el}(L_\text{e}, r) /k_\text{B} T \right] & \mbox{if $\exp\left[ -U_\text{el}(L_\text{e}, r) / k_\text{B} T \right] > \alpha$},  \\
    0 & \mbox{otherwise}.
  \end{array} 
\right.
\label{seff}
\end{equation}
$S_\text{eff}$ counts the area where adsorption probability is higher than $\alpha$. For $\alpha \to 0+$ the original $S_\text{m}=2\pi a^2$ is reproduced. Here, $\alpha=0.02$ was used to get the best fit for the numerical data. As it has been shown in Tab.\ref{tab:ratio_le}, the above analytic approximation matches the data within $10\%$ error margin.
\begin{table}[htb]
\centering
\begin{tabular}{|c|c|c|c|c|c|}
\hline
              &                            &                                        &  correlation & effective size & effective size  \\
  $L_\text{e}$  &  $\uptheta_\text{max}$  & $\Delta \uptheta_\text{max}$    &  coefficient &  (numerical)   & (analytic Eq.(\ref{seff})  \\
             &                            &                                        &                    &                      &  $\alpha=0.02$)   \\
\hline
0.0 & 0.547     & 0.002      &  ---       & 6.28 & 6.28     \\
0.1 & 0.51008 & 0.00005  & 0.9994 &  6.76  & 6.92   \\
0.2 & 0.48535 & 0.00014  & 0.9975 &  7.08  & 7.48   \\
0.5 & 0.43538 & 0.00017  & 0.9983 &  7.88  & 9.00   \\
1.0 & 0.37185 & 0.00020  & 0.9987 &  9.24  & 11.24 \\
2.0 & 0.27660 & 0.00019  & 0.9990 &  12.44 & 14.92 \\
5.0 & 0.14623 & 0.00015  & 0.9968 &  23.48 & 21.76 \\
\hline
\end{tabular}
\caption{Maximal random coverage ratio $\uptheta_\text{max}$, its standard deviation $\Delta \uptheta_\text{max}$ and effective molecule area for different values of $L_\text{e}$ after applying the Feder's law correction. Correlation coefficient describes agreement between numerical data and relation (\ref{federslaw}). $1$ or $-1$ means perfect fit. Values of $L_\text{e}$ and effective sizes are expressed in units of $a$ and $a^2$, respectively.}
\label{tab:ratio_le}
\end{table}
\subsubsection{Autocorrelations}
\begin{figure}[hbt]
  \vspace{1cm}
  \centering
  \includegraphics[width=2.5in]{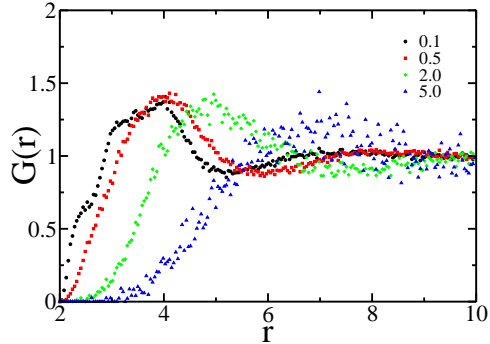}
  \caption{Spatial autocorrelation function. It was obtained for a collector size $200 a$ x $200 a$ with different interaction range $L_\text{e}$. All functions are normalised to be equal to $1$ at infinity. The $r$ is expressed in units of $a$.}
  \label{fig:autocor}
\end{figure}

Autocorrelation function $G(r)$, also known as two point correlation function, is defined as a mean density of adsorbed dimers at a given distance from the centre of one adsorbed molecule. Here, the density is normalised to the overall mean density in a covering layer. Therefore, the autocorrelation function presented in Figure \ref{fig:autocor} approaches  $1.0$ with growing distance.
Autocorrelation behaviour is typical. For small electrostatic interaction range, the rapid grow starts at $r=2a$. As expected, an increase in $L_\text{e}$ makes this grow slower. For $L_\text{e}=5.0a$, there are almost no neighbours in the distance closer than $r=3a$. The maximum of $G(r)$ corresponds to the most probable distance between molecules. With the grow of $L_\text{e}$ it moves to the right and becomes wider, but only for high $L_\text{e}$ values. The data are presented in Table~\ref{tab:distances}.
\begin{table}[htb]
\centering
\begin{tabular}{|c|c|c|c|}
\hline
$L_\text{e}$ & characteristic distance & dispersion \\
\hline
0.1 & 3.68 & 1.215  \\
0.2 & 3.80 & 1.134  \\
0.5 & 4.00 & 1.215  \\
1.0 & 4.29 & 1.296  \\
2.0 & 4.98 & 1.215  \\
5.0 & 6.84 & 1.539  \\
\hline
\end{tabular}
\caption{Characteristic distance between neighbouring dimers and it's dispersion (FWHM) dependence on interaction range $L_\text{e}$. All values are expressed in units of $a$.}
\label{tab:distances}
\end{table}
For $L_\text{e} < 1.0a$, the maximum is followed by the minimum, which corresponds to the excluded volume effect around nearest neighbours. Careful reader may also observe here second smaller maximum. 
\par
The phenomenological approach shows that autocorrelation function for spheres is characterised by a log singularity when ($r \to 2a^{+}$) and superexponential decay for $r \to \infty$ \cite{bib:Swendsen1981, bib:Torquato2006}. These results were also confirmed for non-interacting dimers \cite{bib:Ciesla2012}. Here, screened electrostatic potential makes autocorrelations go to $0$ when  $r \to 2a^{+}$. Moreover, the long range limit is also affected because the electrostatic interaction, obviously, vanishes not as fast as hard-core interaction. Therefore in this case, the superexponential decay cannot be expected.
\subsubsection{Order parameter}
\begin{figure}[hbt]
  \vspace{1cm}
  \centering
  \includegraphics[width=2.5in]{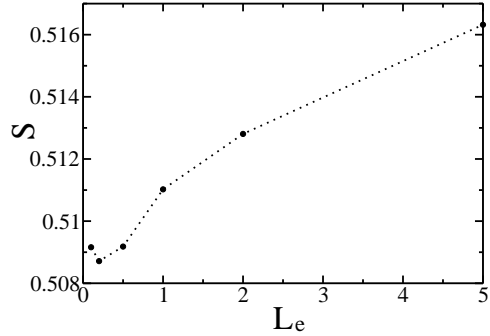}
  \caption{Order parameter for different interaction range. The presented value is a mean taken from 20 to 100 simulations depending on collector size. The time of single simulation was $10^5$.The $L_\text{e}$ is expressed in units of $a$.}
  \label{fig:order_le}
\end{figure}

Dimers have a non-uniform shape and therefore some orientational ordering may occur in adsorbed layers. This phenomenon is important from a practical point of view although it affects important mechanical, electrochemical and optical properties of a formed layer.
To determine whether any orientational order appears in a coverage, measurement of such an ordering is needed. It can be based on the function $S(\alpha)$, which is the sum of squared scalar products between molecules orientation and a given angle $\alpha$ \cite{bib:Ciesla2012}:
\begin{figure}[hbt]
  \vspace{1cm}
  \centering
  \includegraphics[width=2.5in]{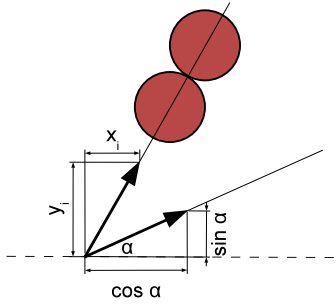}
  \caption{Calculation of $S(\alpha)$ function (see Eq.(\ref{sa})) for a given direction $\alpha$. Both vectors are assumed to have unit length.}
  \label{fig:angle}
\end{figure}
\begin{equation}
S(\alpha) = \frac{1}{N} \sum_{i=1}^N \left(x_i \cos \alpha + y_i  \sin \alpha \right)^2,
\label{sa}
\end{equation}
where $(x_i, y_i)$ denotes a unit vector along the direction of $i$-th molecule in a layer (see Fig.\ref{fig:angle}) and $N$ is the total number of adsorbed dimers. In the case of all particles oriented along one direction, $S(\alpha)$ will have a maximum (equal to $1$) for this specific direction and a minimum (equal to $0$) for perpendicular direction. When there is no orientational order inside adsorbed layers, the $S(\alpha)$ will be constant and equal to $0.5$ for all directions. Therefore, mean particles orientation is given by the maximum of $S(\alpha)$ function and can be estimated from:
\begin{equation}
\tan 2\alpha_\text{max} = \frac{\sum_{i=1}^N x_i \, y_i}{
\sum_{i=1}^N x_i^2 - \sum_{i=1}^N y_i^2},
\end{equation}
and the order parameter can be defined as $S \equiv S(\alpha_\text{max})$. It is worth to notice that the above equation is satisfied by both $\alpha_\text{max}$, and $\alpha_\text{max} + \pi / 2$; so it should be determined which one is a maximum and which is a minimum. Dependence of $S$ on interaction range $L_\text{e}$ is shown in Fig.\ref{fig:order_le}.
All the values are close to $S=0.5$, which means the orientations of particles are randomly distributed and there is no significant global ordering. However, there is small but noticeable growth of $S$ with the growth of interaction range. We have checked that this effect is more evident for smaller collectors sizes. It is probably induced by adsorption conditions near collector boundaries. With the grow of interaction range the density of adsorbed molecules decreases and the influence of boundaries propagates on larger distances, which explains the growth of global ordering.
\par
\begin{figure}[hbt]
  \vspace{1cm}
  \centering
  \includegraphics[width=2.5in]{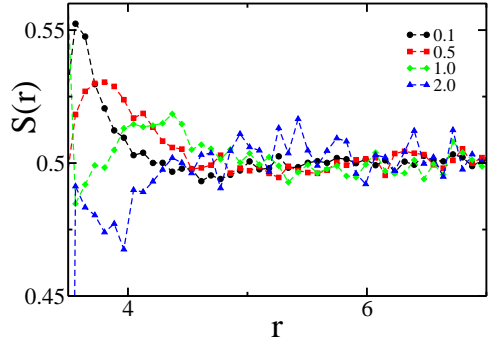}
  \caption{Order autocorrelation function for different Debye screening lengths $L_\text{e}$. The $r$ is expressed in units of $a$.}
  \label{fig:order_cor}
\end{figure}

To complete this picture, we measured also two-point correlation functions of the orientational order. It allows the measurement of local orientational ordering. Results in Fig. \ref{fig:order_cor} present a mean value of the scalar product of main particle axes separated by a given distance.
When interaction range is small ($L_\text{e} \le 0.5a$) the nearest neighbours slightly prefer parallel alignment; whereas when $L_\text{e} \ge 1.0a$, the $S(r)$ drops below $0.5$ which means slight tendency to perpendicular alignment. For larger distances, all plots quite quickly approach $0.5$, which is accordance with previous observations and means lack of global ordering.
\subsection{Adsorption kinetics}
The growth of covering layer for homogeneous systems can be described by one dimensional differential equation:
\begin{equation}
\frac{d}{dt}\uptheta(t) = k \, \text{ASF}(\uptheta) \, c(0,t),
\label{kineticstheta}
\end{equation}
where $\text{ASF}(\uptheta)$, known also as Available Surface Function, is a ratio of uncovered adsorption-ready surface to the whole collector area; $k$ is an adsorption reaction constant; and $c(0, t)$ is a molecules concentration near the collector surface. 
\begin{figure}[hbt]
  \centering
  \includegraphics[width=4in]{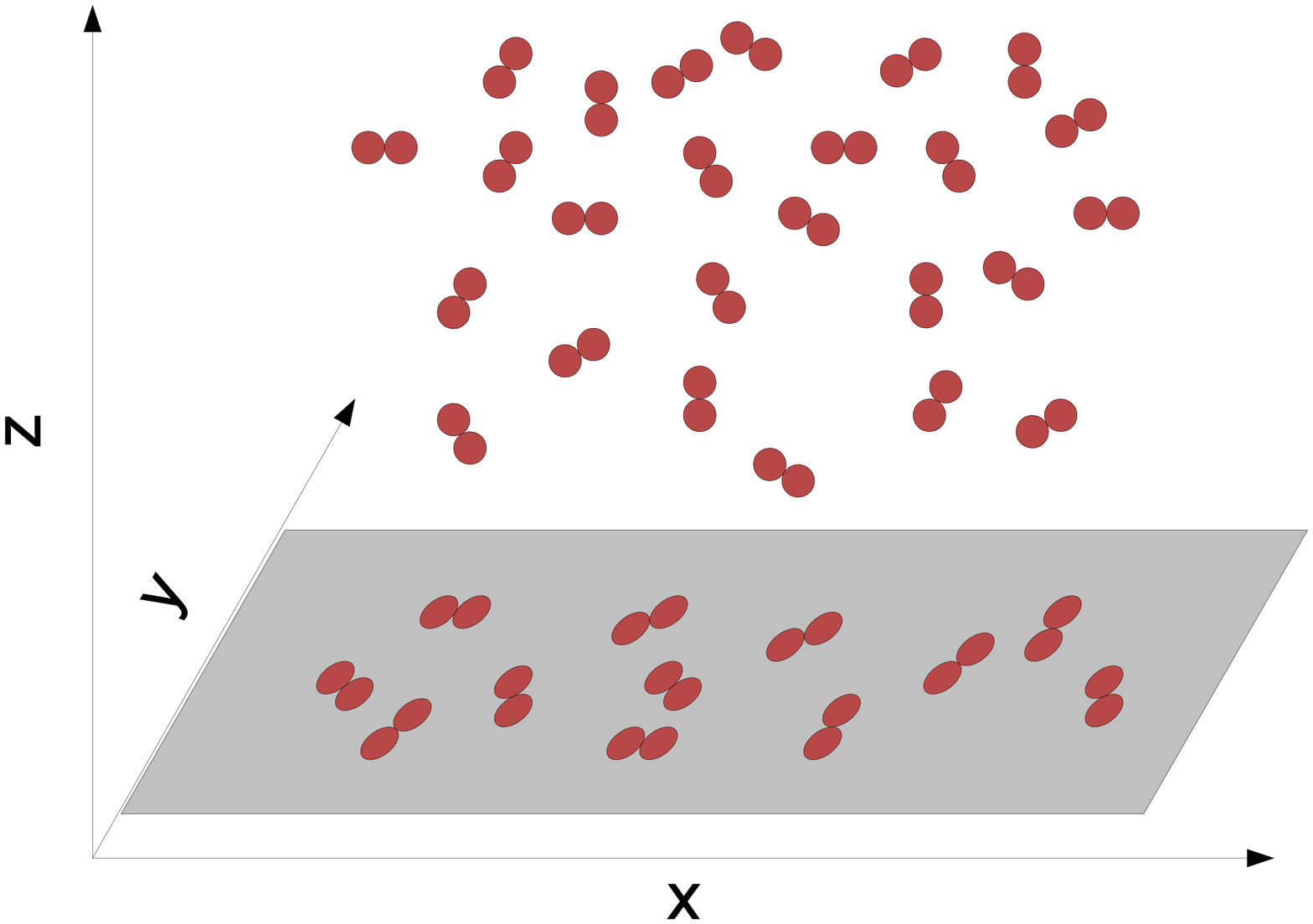}
  \caption{Snapshot from experiment. Particles are transported along $z$-axis from a bulk to surface proximity.}
  \label{fig:diffusion}
\end{figure}
The RSA kinetics assumes that the only factor affecting the speed of adsorption layer growth is decreasing probability of finding large enough, unallocated place on the collector surface. From a physical point of view, it is equivalent to the assumption that concentration $c(0, t)$ of molecules in a solution near the layer  is constant. However, the value of $c(0, t)$ is generally a result of different physical processes which move particles from a bulk to a surface neighbourhood (see Fig.\ref{fig:diffusion}). Their details depend on an experiment preparation and its environment. On the other hand, the most common process of transport in the majority of experiments is diffusion:
\begin{equation}
\frac{\partial c(z, t)}{\partial t}  = D \frac{\partial^2 c(z, t)}{\partial z^2},
\label{diffusion}
\end{equation}
where $D$ is a diffusion constant. Typically, the number of adsorbed particles is negligible in comparison to the total number of particles in bulk. Therefore, far from the surface, concentration of particles remains constant:
\begin{equation}
\lim_{z \to \infty} c(z,t) = c_{\infty},
\label{b1}
\end{equation}
where $c_{\infty}$ is a bulk concentration of particles. On the other hand, the only mechanism that removes particle from a bulk is adsorption and the process occurs at the surface ($z=0$). Therefore, the continuity of particles flux at the surface together with Eq.(\ref{kineticstheta}) follows to:
\begin{equation}
-D \frac{\partial c(0,t)}{\partial z} = k \, \text{ASF}(\uptheta) \, c(0, t)
\label{b2}
\end{equation}
Equations (\ref{b1}) and (\ref{b2}) are known as mixed boundary conditions or Robin boundary conditions. Due to nonlinearity of the $\text{ASF}(\uptheta)$, the equation set (\ref{kineticstheta})--(\ref{b2}) can be solved only numerically. However, the first step to determine real time adsorption kinetics $\uptheta (t)$ is to determine Available Surface Function $\text{ASF}(\uptheta)$.
\subsubsection{Measurement of the $\text{ASF}(\uptheta)$}
\label{sec:asf}
\begin{figure}[hbt]
  \vspace{1cm}
  \centering
  \includegraphics[width=2.5in]{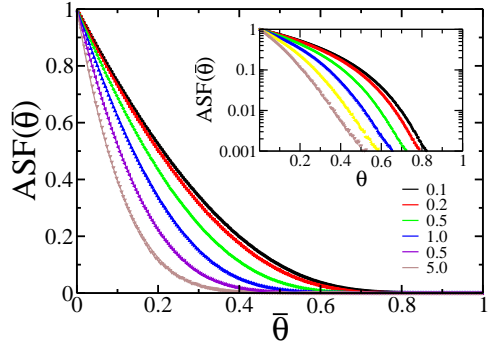}
  \caption{Available Surface Function for different range of electrostatic interaction. Inset shows the same data in a log scale. Solid lines correspond to fit given by Eq.(\ref{asf_fit}).}
  \label{fig:asf}
\end{figure}

Available Surface Function can be measured by estimating the RSA success rate of placing a particle on a collector for a given coverage. The total number of attempts was set typically as $10^3$; however, if all of the attempts failed the simulation was performed till the first successful attempt. Obtained results are presented in Figure \ref{fig:asf}.
The $\text{ASF}(\uptheta)$ behaviour is typical. For low coverages $\text{ASF}(\uptheta)$ can be approximated by a linear or square function of $\uptheta$. When $\uptheta$ approaches $\uptheta_{max}$, the exponential decrease is observed. The typical analitical approximation of $\text{ASF}(\uptheta)$ for elongated particles like ellipsoids and spherocylinders is \cite{bib:Ricci1992} $\text{ASF}(\uptheta) = (1 + a_1 \uptheta + a_2 \uptheta^2 + a_3 \uptheta^3)(1-\bar{\uptheta})^4$. However, the term $(1-\uptheta)^4$  in the $\text{ASF}(\uptheta)$ is clearly related to the existence of a long-time kinetics governed by a modified Feder's law: $\uptheta_{max}-\uptheta(t) = A t^{-1/3}$. Therefore, to be consistent with Eq.(\ref{federslaw}) for $d=2$, which has been confirmed by results presented in Fig.\ref{fig:feder}, here we proposed the following approximation:
\begin{equation}
\label{asf_fit}
ASF(\bar{\uptheta}) = (1 + a_1 \bar{\uptheta} + a_2 \bar{\uptheta}^2 + a_3 \bar{\uptheta}^3)(1-\bar{\uptheta})^3,
\end{equation}
where $\bar{\uptheta} = \uptheta / \uptheta_\text{max}$. The values of fitted parameters are presented in Table \ref{tab:asf_fit}. 
\begin{table}[htb]
\centering
\begin{tabular}{|c|c|c|c|c|}
\hline
$L_\text{e}$ & $a_1$ & $a_2$ & $a_3$ & correlation coefficient \\
\hline
0.1 &   0.2971 & -0.4840 & -1.8782 & 0.999996 \\
0.2 &   0.1935 & -1.1530 & -1.2577 & 0.999990 \\
0.5 & -0.2060 & -3.2556 &   1.9832 & 0.999975 \\
1.0 & -1.1151 & -4.0851 &   5.3544 & 0.999987 \\
2.0 & -2.9237 &   0.0330 &   3.5859 & 0.999888 \\
5.0 & -5.4794 &   9.5474 & -5.1290 & 0.999850 \\
\hline
\end{tabular}
\caption{Fitted parameters according to Eq.(\ref{asf_fit}).}
\label{tab:asf_fit}
\end{table}
Relation (\ref{asf_fit}) gives the best fit when interaction range is small. For large $L_\text{e}$ the ASF quickly approaches exponential function rather than polynomial one, which can be clearly seen when using logarithmic scale. In general, the fit given by (\ref{asf_fit}) breaks when $ASF$ becomes exponential. For $L_\text{e}=0.1a$, it takes place at $\bar{\uptheta} \approx 0.7$, whereas for $L_\text{e}=5.0a$ at $\bar{\uptheta} \approx 0.3$.
\subsubsection{Real time kinetics}
\begin{figure}[hbt]
  \vspace{1cm}
  \centering
  \includegraphics[width=2.5in]{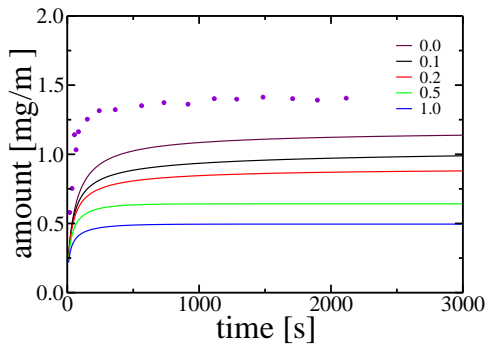}
  \caption{Real time kinetics of dimers adsorption. Different lines corresponds to different $L_e$ values. Dots are taken from experiment \cite{bib:Mollmann2005}.}
  \label{fig:kinetics}
\end{figure}

Numerical results of dimers adsorption were compared with experimental data of insulin adsorption obtained by Mollmann et al.\,\cite{bib:Mollmann2005}. Therefore, we set the diffusion coefficient  $D=100$ \textmu m\textsuperscript{2}/s, and dimer surface $S_\text{m} = 8.75 \times 10^{-6}$ \textmu m\textsuperscript{2}. The bulk concentration was $10^{-2}$ mg/ml. The value of adsorption reaction rate had not been specified explicitly; therefore, we used $k=1$ \textmu m/s in our calculation. Results are presented in Figure \ref{fig:kinetics} were obtained by solving Eqs. (\ref{kineticstheta})--(\ref{b2}) using $\text{ASF}(\uptheta)$ designated in Sec.\,\,\ref{sec:asf}. The details about numerical procedure used here, to solve diffusion equation are described in \ref{kinetics_procedure}. The units used here are commonly used by experimenters.
Presented plots drawn for different electrostatic interaction range are similar. The most noticeable difference is connected with maximum coverage ratio $\uptheta_\text{max}$, which strongly depends on interaction range (see Fig.\ref{fig:ratio_le} and Tab.\ref{tab:ratio_le}). However, it can be noticed that the saturation occurs faster with the growth of interaction range, which is a direct consequence of the $\text{ASF}(\uptheta)$ shape. The difference between simulation and experimental data result mainly from different values of the maximal random coverage ratios. The causes of this effect were already discussed earlier \cite{bib:Ciesla2012}. Apart from this, the shapes of numerically and experimentally determined kinetics are similar. 
\section{Summary}
The random maximal coverages ratio for electrostatically interacting di\-mers was measured using the RSA method. Repulsive interactions lowers the maximum possible coverages ratio the more the larger interaction range $L_e$ is, eg. $\uptheta_\text{max} = 0.510$ for $L_\text{e}=0.1a$ and  $\uptheta_\text{max} = 0.146$ for $L_\text{e}=5.0a$; however, it does not affect the validity of the Feder's law. The measurement of autocorrelation function allows to find out the characteristic distance between neighbouring molecules and its value changes from $3.68$ for $L_\text{e}=0.1a$ to $6.84$ for $L_\text{e}=5.0a$. The spontaneous orientational ordering is imperceptible at the global scale but it can occur locally and can be exaggerated by the presence of boundaries. The adsorption kinetics depends mainly on maximal coverage ratio but it can be observed that saturation value is approached faster for larger $L_\text{e}$. 
\par
This work was supported by MNiSW/N N204 439040.
\appendix
\section{Evaluating the kinetics of adsorption}
\label{kinetics_procedure}
There are known some algorithms for calculating adsorption kinetics in case of adsorption \cite{ bib:Adamczyk1987, bib:Erban2007}; however, we want to introduce here another scheme based only on diffusion equations (\ref{kineticstheta})--(\ref{b2}), which can be easily applied to any chemical reaction. 
Therefore, the diffusion equation is solved using standard Crank-Nicholson method; however, the introduction of boundary conditions requires additional operations after each step of the algorithm, which are described below:
\begin{description}
\item[i)] the concentration having the largest $z$ coordinate is set to $c_{\infty}$.
\item[ii)] the adsorption probability is given by:
\begin{equation}
p(t) = k \, \text{ASF}(\uptheta) \, \Delta t / \Delta z.
\end{equation}
It has been proved that the above relation fully satisfies Robin boundary conditions (\ref{b2}) \cite{bib:Erban2007_2}. Following relations are direct consequences of the above. The number of particles near the surface is equal to 
\begin{equation}
n(t) = c(0, t) \, S_\text{c} \, \Delta z.
\end{equation}
Thus, the number of adsorbed molecules is:
\begin{equation}
n_\text{A}(t) = n(t) \, p(t) = c(0, t) \,\, S_\text{c} \, k \,\, ASF(\uptheta) \, \Delta t .
\end{equation}
$\Delta x$ and $\Delta t$ are space and time discretisation steps used in Crank-Nicholson algorithm. The coverage increase corresponding to the above number $n_\text{A}$ is given by:
\begin{equation}
\Delta \uptheta(t) = n_\text{A}(t) \, S_\text{m}/S_\text{c} = c(0, t) \, k \, \text{ASF}(\uptheta) \,  \Delta t \, S_\text{m}
\end{equation}
so the concentration near surface changes according to 
\begin{equation}
c(0, t+\Delta t) = c(0, t) - n_\text{A}(t) / (S_\text{c} \, \Delta z) = c(0, t) [1 - k \, \text{ASF}(\uptheta) \, \Delta t / \Delta z]
\end{equation}
\end{description}
To use the above algorithm, the diffusion constant $D$, Available Surface Function $\text{ASF}(\uptheta)$ and reaction rate $k$ is needed. The diffusion constant depends on adsorbed molecule properties (shape, mass) as well as solvent properties and it can be measured experimentally. $\text{ASF}(\uptheta)$ can be easily determined from RSA simulation as a RSA success rate of placing a particle on a given layer. The adsorption reaction rate $k$ can be expressed in terms of physical parameters characterising the system such as the particle diffusion coefficient, specific energy distribution, depth equilibrium state, and height of the adsorption energy barrier \cite{bib:Adamczyk-book, bib:Adamczyk2000}. For barrier-less adsorption regime, the adsorption constant is given by 
\begin{equation}
k =\frac{D}{\delta_\text{a} \left(\frac{1}{2} + \ln \frac{\delta_\text{a}}{\delta_\text{m}} \right)},
\end{equation}
where $\delta_\text{a}$ is the thickens of adsorption boundary and $\delta_\text{m}$ is the minimum distance between molecule in bulk phase and surface \cite{bib:Adamczyk2000}. Reaction rate can also be measured experimentally or used as a parameter which can be fitted to obtain the best possible agreement between experiment and numerical simulation.

\end{document}